\documentclass[12pt,noshowpacs,nofootinbib,notitlepage,amsmath]{revtex4-1}
\allowdisplaybreaks
\usepackage{graphicx,color}
\usepackage[colorlinks=true,citecolor=blue,linkcolor=blue,urlcolor=blue]{hyperref}
\usepackage[charter]{mathdesign}
\DeclareSymbolFontAlphabet{\mathcal}{symbols}
\DeclareSymbolFont{symbols}{OMS}{xmdcmsy}{m}{n}
\DeclareSymbolFont{largesymbols}{OMX}{xmdcmex}{m}{n}
\SetSymbolFont{symbols}{bold}{OMS}{xmdcmsy}{b}{n}

\begin{document}  
\title{\color{blue}\Large The Early Universe with High-Scale Supersymmetry}
\author{Sibo Zheng}
\email{sibozheng.zju@gmail.com}
\affiliation{Department of Physics, Chongqing University, Chongqing 401331, P.R. China}
\begin{abstract}
A small tensor-to-scalar ratio $r$ may lead to distinctive phenomenology of high-scale supersymmetry.
Assuming the same origin of SUSY breaking between the inflation and visible sector,
we show model independent features.
The simplest hybrid inflation, together with a new linear term for the inflaton field  
which is induced by large gravitino mass, 
is shown to be consistent with all experimental data for $r$ of order $10^{-5}$. 
For superpartner masses far above the weak scale we find that the reheating temperature after inflation 
might be beneath the value required by thermal leptogenesis 
if the inflaton decays to its products perturbatively,
but above it if non-perturbatively instead.
Remarkably, the gravitino overproduction can be evaded in such high-scale supersymmetry because of the kinematically blocking effect.

\end{abstract}
\maketitle

\section{Introduction }
\label{intro}
After the discovery \cite{Higgs} of standard model (SM) Higgs boson at the Large Hadron collider (LHC),
low-scale supersymmetry (SUSY) \cite{susy} which is favored 
by the naturalness argument \cite{naturalness} has been extensively explored.
These studies show the difficulties in the both theoretic explanation of 125 GeV Higgs mass and experimental fits to the LHC data. 
The second run of LHC will shed light on the prospect of such natural SUSY models.
Given above consideration some efforts have been devoted to the study of high-scale SUSY.

Even though high-scale SUSY cannot be detected at the $14$ TeV LHC,
they can be still studied via their effects on the evolution of early universe.
Measurement on the tensor-to-scalar ratio $r$ via experiments 
such as WAMP, Plank and BICEP that are devoted to measure Cosmic Microwave Background (CMB) temperature anisotropy and polarization during inflation,
may probe high-scale SUSY with mass spectrum far above the weak scale.
The measured value of $r$ reported by the Plank Collaboration is of order $r< 0.11$ 
at 95 \% CL \cite{1502.02114, 1303.5076,1303.5082},
from which the energy scale of inflation can be directly inferred.
Since the energy scale of inflation is proportional to $r^{1/4}$, 
it is only mildly sensitive to $r$.
So the study of high-scale SUSY remains well motivated as long as $r$ is not extremely small.

In this paper, we consider inflationary models with $r$ far below the Plank bound value $r_{c}=0.11$.
The motivation is mainly based on two facts.
At first, the stability of the SM electroweak vacuum requires $H\leq 0.04 ~h_{\text{max}}$ \cite{1505.04825}, 
where $h_{\text{max}}$ refers to the value $h$ at which the Higgs potential is maximal. 
For the central value of top quark pole mass
$h_{\text{max}}\sim 10^{10}$ GeV, 
which implies that $H$ should be smaller than Plank bound value $H_{\text{c}}\sim10^{16}$ GeV corresponding to $r_c$ \cite{1504.08093}.
In this sense small $r<<r_{c}$ is more favored to guarantee the electroweak vacuum stable against quantum fluctuation during inflationary epoch.
Secondly, for $r<<r_{c}$, it can still generate SUSY mass spectrum large enough to escape the LHC constraints.

For simplicity we adopt the assumption that the inflation and visible (namely the minimal supersymmetric standard model (MSSM )) sector share the same origin of SUSY breaking. 
This assumption is rational, as it can be realized in model building.
Moreover, it allows us to discuss reheating in the early universe after inflation,
once the SUSY mass spectrum and the inflaton decay are identified explicitly.

The paper is organized as follows.
In section 2 we re-analyze the model independent consequences from above assumption within the range $r<<r_{c}$.
In section 3, we consider hybrid inflation as an example in the course of high-scale SUSY breaking 
\footnote {For earlier attempts to address this issue, see, .e.g, \cite{1009.5340, 1211.0994}.}.
We will show that a new linear term for inflaton field 
with a large coefficient proportional to $m_{3/2}$ 
affects the inflation significantly,
and the simplest hybrid inflation is consistent with $r$ of order $10^{-5}$.

In the second part of this paper we discuss the reheating in the early universe after inflation in section 4. 
In particular, reheating temperature $T_R$ after inflation is estimated
for superpartner mass spectrum $m_{0}$ above $\mathcal{O}(100)$ TeV.
We find that $T_R$ might be beneath the value $\sim 10^{9}$ GeV required by thermal leptogenesis
if inflaton decays to its products perturbatively,
but above it if non-perturbatively instead.
The gravitino overproduction in conventional high-scale SUSY can be easily evaded 
because of kinematically blocking effect.
Finally we conclude in section 5.

\section{Implications of the value of $r$ to Inflation}
In this section we revise the model independent implications (together with Plank and 9-year WAMP data) to single-field inflation for $r<< r_{c}$.
These results provide useful information on the model building of inflation,
as a reliable inflation model should at least explain observable quantities as what follows.

$(1)$, First of all the scale of energy density during inflation is directly related to $r$ as,
\begin{eqnarray}{\label{energydensity1}}
V^{1/4}=\left(24\pi^{2}M^{4}_{P}\epsilon A_{s}\right)^{1/4}.
\end{eqnarray}
where $A_s$ is the amplitude of the power spectrum of the curvature perturbation and $\epsilon=r/16$ in the context of single-field inflation.
Recall that $A_{s}^{1/2}=H^{2}(\phi_{*})/2\pi\dot{\phi}$,
where $\phi_{*}$ is the value of $\phi$ when wavenumber $k_{*}=0.05$ Mpc$^{-1}$ crossed outside the horizon.
Substituting the reported value $A^{1/2}_{S}\simeq 3.089\times 10^{-5}$ by Plank Collaboration \cite{1303.5076} into Eq.(\ref{energydensity1}) gives rises to,
\begin{eqnarray}{\label{energydensity2}}
V^{1/4}\simeq 2\times 10^{16} \cdot \left(\frac{r}{0.20}\right)^{1/4} \text{GeV}.
\end{eqnarray}
Eq.(\ref{energydensity2}) is valid independent of the explicit form of inflation potential.
If the MSSM and inflation share the same origin of SUSY breaking, 
as we have assumed in this paper, 
the high-scale SUSY breaking scale $\sqrt{F}$ in particle physics will be order $\simeq V^{1/4}$.
For example, $\sqrt{F} $ is of order $\sim 10^{16}$ GeV for $r\sim 0.1$,
and slightly reduced to be of order $\sim 10^{15}$ GeV for $r\sim 10^{-5}$.

$(2)$, In the context of slow roll inflation 
the spectral index $n_s$ (for scalar) and  $n_t$ (for tensor) are given by, 
\begin{eqnarray}{\label{index}}
n_{s}-1\simeq 2\eta-6\epsilon,~~~~~~n_{t}\simeq -2\epsilon,
\end{eqnarray}
respectively.
Here $\epsilon=\frac{M^{2}_{P}}{2}\left(V_{,\phi}/V\right)^{2}$ and $\eta=M^{2}_{P}V_{,\phi\phi}/V$, 
with subscript denoting derivative of $V$ over $\phi$.
The combination of Plank and 9-year WAMP data measures the value of $n_s$ in high precision \cite{1303.5082}, 
\begin{eqnarray}{\label{ns}}
n_{s}= 0.9603 \pm 0.0073,
\end{eqnarray}
For $r<< r_{c}$ one finds that $\eta\simeq -0.02$.
This tight bound is crucial to constrain inflation model.

$(3)$, The gravitino mass $m_{3/2}$ can be determined.
The constant superpotential $W_{0}=m_{3/2}M^{2}_{P}$, 
which is required to cancel out positive $F^{2}$ term in the potential so as to explain the smallness of cc,  gives rise to 
\begin{eqnarray}{\label{gravitinomass}}
m_{3/2}=\frac{F}{\sqrt{3}M_{P}}
\end{eqnarray}

$(4)$, Finally the number of e-fold that $k_{*}$ undergoes during inflation is given by,
\begin{eqnarray}{\label{N}}
N\simeq \int^{\phi_{in}}_{\phi_{end}}\frac{d\phi}{M^{2}_{P}}\frac{V(\phi)}{V_{,}(\phi)}\simeq \int^{x_{in}}_{x_{end}}\frac{dx}{\sqrt{2\epsilon(x)}},
\end{eqnarray}
where $x=\phi/M_{P}$. Subscript ``in" and ``end" corresponds to initial and end value of $x$ during inflation, respectively.
For realistic inflation models, $N$ is bounded as $50 \leq N\leq 60$.
If $\epsilon$ doesn't change significantly during inflation, 
Eq.(\ref{N}) can be expressed as $\Delta\phi/M_{P}\simeq \sqrt{2\epsilon} N\simeq \sqrt{\frac{r}{8}} N$.
This is known as Lyth bound \cite{Lythbound},
which shows the need of small field inflation for $r<< r_{c}$.

\section{The Simplest Hybrid Inflation}
The section is devoted to the study of inflation building in the course of high-scale SUSY.
We take the simplest hybrid inflation as an explicit illustration.
We will show that a new linear term due to the assumption adopted in this paper significantly affects the choice on initial condition. 
Also this assumption introduces new constraints on parameters in the model, 
which make the simplest hybrid inflation only possible with $r$ of order $10^{-5}$.

\subsection{Scalar Potential}
The scalar potential in hybrid inflation is constructed from superpotential $W$, 
\begin{eqnarray}{\label{W}}
W=\kappa \Phi(\bar{\Psi}\Psi-M^{2}),
\end{eqnarray}
and Kahler potential $K$,
\begin{eqnarray}{\label{kahler}}
K=\mid\Phi\mid^{2}+\mid\Psi\mid^{2}+\mid\bar{\Psi}\mid^{2}+k_{1}\frac{\mid\Phi\mid^{4}}{M^{2}_{P}}.
\end{eqnarray}
Here $\Phi$ denotes the inflaton superfield, with its lowest component inflaton field $\phi$.
$\Phi$ is a singlet of standard model gauge groups $G=SU(3)_{c}\times SU(2)_{L}\times U(1)_{Y}$.
$\Psi$ and $\bar{\Psi}$ denote waterfall superfields which are in the bi-fundamental representation of $G$ \footnote{
Alternatively, $G$ can be extended to include a local $U(1)_{B-L}$ symmetry so as to explain leptogenesis.}.
The Kahler potential in Eq.(\ref{kahler}) takes into account the non-canonical term,
with $k_1$ a real coefficient.
The non-canonical $k_1$ term provides the inflaton mass term.
$M_P=2.4\times 10^{18} $ GeV is the reduced Plank mass,
while $M$ is assumed to be far below $M_{P}$.

Substituting Eq.(\ref{W}) and Eq.(\ref{kahler}) into the SUGRA potential 
\begin{eqnarray}{\label{SUGRA}}
V=e^{K/M^{2}_{P}}\left[K^{i\bar{j}}D_{i}WD_{\bar{j}}W-3\frac{\mid W\mid^{2}}{M^{2}_{P}}\right],
\end{eqnarray}
one obtains the scalar potential of hybrid inflation.
Here $K^{i\bar{j}}=K^{-1}_{i\bar{j}}$ is the Kahler metric and $D_{i}W=\partial_{i}W+K_{i}W/M^{2}_{P}$.
It is well known that the history of inflation can be naturally divided into two periods. 
In the first one inflation usually starts from an initial value of order $M_{P}$ towards to $\phi_c$ 
which is a critical value separating the two periods.
The vacuum in the first period corresponds to $\Psi=\bar{\Psi}=0$,
from which the energy density reads from Eq.(\ref{SUGRA}) \cite{1007.5152},
\begin{eqnarray}{\label{density}}
V=\frac{1}{2}m_{\phi}^{2}\phi^{2}+\kappa^{2}M^{4}\left[1+\gamma \frac{\phi^{4}}{8M^{4}_{P}}+\frac{\kappa^{2}}{16\pi^{2}}\ln\frac{\kappa^{2}\phi^{2}}{2\Lambda^{2}}\right]
+2\sqrt{2}\kappa M^{2}m_{3/2} \phi\cos\theta,
\end{eqnarray}
where we have defined $\Phi=\phi e^{i\theta}/\sqrt{2}$.
Here $\gamma=1-7k_{1}/2+2k^{2}_{1}$ and inflaton mass $m_{\phi}$ is given by for negative $k_1$,
\begin{eqnarray}{\label{mass}}
m_{\phi}=\sqrt{-k_{1}}\kappa M^{2}/M_{P}.
\end{eqnarray}
The $\log$-term in Eq.(\ref{density}) represents the contribution due to mass splitting in waterfall fields \cite{9406319}, with $\Lambda$ the cut-off scale.
The linear term with coefficient proportional to $m_{3/2}$ arises from a constant superpotential  $W_{0}=m_{3/2}M^{2}_{P}$ added to $W$,
which is needed to cancel out positive contribution to energy density due to SUSY breaking,
and explain the smallness of cc.

When $\phi$ approaches to $\phi_{c}=\sqrt{2}M$, 
$\phi$ becomes massless as shown from its mass squared $m^{2}_{\Psi}=-4\kappa^{2}M^{2}+2\kappa^{2}\phi^{2}$.
After the time when $\phi$ is below $\phi_c$,
$\Psi$ starts to roll towards to its global minimum value $\Psi=M$ from $\Psi=0$,
which is known as the second period of inflation.
The evaluation of field $\phi$ (including  angular component $\theta$) and $\Psi$ during each period is determined by their equations of motion,
\begin{eqnarray}{\label{equationofmotion}}
3H\dot{\phi}&\simeq&m^{2}_{\phi}\phi+2\sqrt{2}\kappa M^{2}m_{3/2}\cos\theta,\nonumber\\
3H\dot{\theta}&\simeq&2\sqrt{2}\kappa M^{2}m_{3/2}\frac{\sin\theta}{\phi},\\
3H\dot{\Psi}&\simeq&-2\kappa^{2}(2M^{2}-\phi^{2})\Psi,\nonumber
\end{eqnarray}
where $H$ the Hubble constant is subject to the Friedmann constraint
\begin{eqnarray}{\label{H}}
H^{2}\simeq 8\pi V/3M^{2}_{P}.
\end{eqnarray}
The time for each period is controlled by the magnitude of $m_{\phi}$ or $\mid m_{\Psi}\mid$ relative to Hubble constant $H$.
As pointed out in \cite{hybrid}, 
the second period is very short in compared with the first one for wide ranges of parameter choices.
Substituting Eq.(\ref{H}) into the last equation in Eq.(\ref{equationofmotion}),
we obtain the constraint for such property,
\begin{eqnarray}{\label{constraints}}
10^{-4}< \kappa< \mathcal{O}(1).
\end{eqnarray}
In the next subsection, we will discuss in more details the initial conditions on inflaton field and the field value of $\phi$ when inflation ends.

\subsection{Initial Conditions}
The inflation usually begins at some field value $\phi_{in}$ near Plank scale.
The choice on $\phi_{in}$ is subtle when the inflaton potential has either a few local minimums at $\phi_{min}$s, or local maximum at $\phi_{max}$s.
If one adopts $\phi_{in}$ bigger than $\phi_{min}$,
inflaton is probably trapped at  these local minimums of inflaton potential along the trajectory,
which leads to inflation with insufficient e-fold number $N\sim 50-60$.
In order to avoid this, one should choose $\phi_{in}<\min\{\phi_{min}\}$.
On the other hand, one wants that inflation proceeds with exactly decreasing $\phi$.
This is only allowed if $\phi_{in}$ is less than $\min\{\phi_{max}\}$.
In other words, we should impose the initial condition 
\begin{eqnarray}{\label{initial}}
\phi_{in}<\min\{\phi_{max},\phi_{min}\}.
\end{eqnarray}

\begin{figure}
\centering
\begin{minipage}[b]{0.7\textwidth}
\centering
\includegraphics[width=4in]{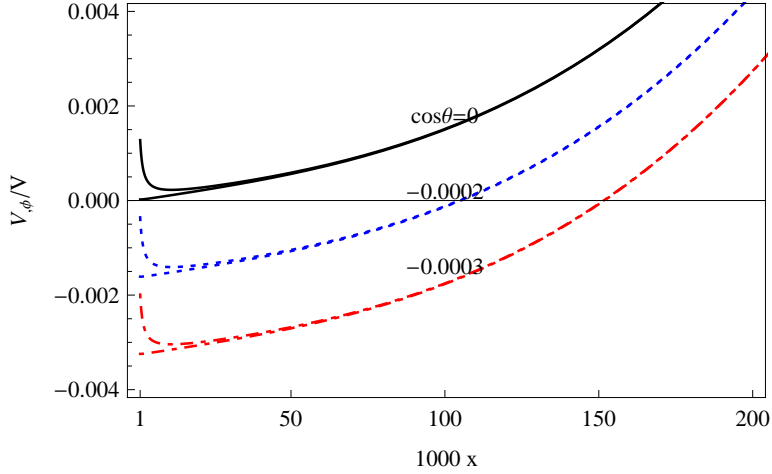}
\end{minipage}%
\caption{Initial condition on $x\equiv\phi/M_{P}$ as function of $\cos\theta$ for $\kappa=0.01, 0.001$.}
\label{extreme}
\end{figure}

In Fig.\ref{extreme} we show how extremes in $V$ depend on $\cos\theta$ and $\kappa$ by evaluating $\sqrt{2\epsilon}=V_{,\phi}/V$.
The sign of $V_{,\phi}/V$ changes when $x\sim 0.10$ for $\cos\theta=-0.002$ and
$x\sim 0.15$ for $\cos\theta=-0.003$.
This implies that $\cos\theta\simeq 0$ for realistic inflation.
Otherwise, $\phi_{in} << M_{P}$, which is too small to provide enough e-fold number $N$.
This observation has been noted in \cite{1007.5152} for $m_{3/2}$ of order electroweak scale,
and further verified for larger value of $m_{3/2}\sim 10^{13}$ GeV.
With initial value $\theta\simeq \pi/2$,
the initial value $\phi_{in}$ can be chosen in wide range, as shown in Fig.\ref{extreme}.
The evaluation of $\phi$ from $\phi_{in}$ is the same as original hybrid model because of absence of linear term in the first equation in Eq.(\ref{equationofmotion}).
In this sense, inflation mainly ends at field value $\phi_{end}=\sqrt{2}M$.

The e-fold number $N$ produced during inflation and $n_s$ can be both 
estimated in terms of slow roll parameters $\epsilon$ and $\eta$ in the model,
which are given by respectively,
\begin{eqnarray}{\label{slowrollpa}}
\eta(x)&\equiv&M^{2}_{P}\frac{V_{,\phi\phi}}{V}
\simeq -k_{1} +\frac{3}{2}x^{2}-\frac{\kappa^{2}}{8\pi^{2}}\frac{1}{x^{2}}, \nonumber\\
\epsilon(x)&\equiv&\frac{M^{2}_{P}}{2}\left(\frac{V_{,\phi}}{V}\right)^{2}
\simeq \frac{1}{2}\left(\sqrt{\frac{8}{3}}\cos\theta-k_{1}x+\frac{1}{2} x^{3}+\frac{\kappa^{2}}{8\pi^{2}}\frac{1}{x}\right)^{2}.
\end{eqnarray}
With $k_{1}\simeq -0.01$ and $\cos\theta\simeq 0$, 
$\eta(x)$ and $\epsilon(x)$ mainly depend on parameter $\kappa$. 

\begin{figure}
\centering
\begin{minipage}[b]{0.8\textwidth}
\centering
\includegraphics[width=4.5in]{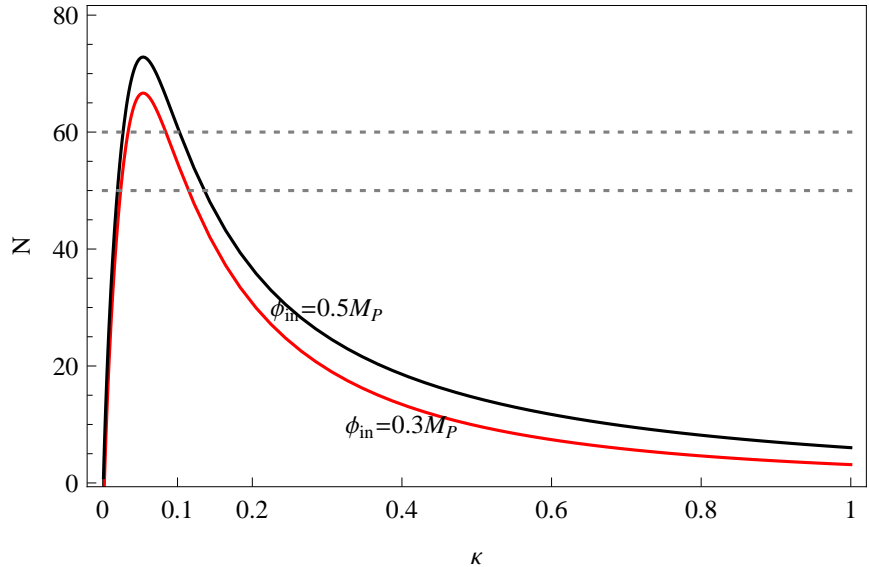}
\end{minipage}%
\caption{$N$ as function of $\kappa$ for initial value $x_{in}=0.3$ (red), 0.5 (black) respectively. Dotted lines represent the uncertainty on experimental value. 
Slow roll condition $\mid\eta\mid <1$ leads to the bound $\phi_{in}\leq M_{P}$. }
\label{e-folds}
\end{figure}

\begin{figure}
\centering
\begin{minipage}[b]{0.8\textwidth}
\centering
\includegraphics[width=4.5in]{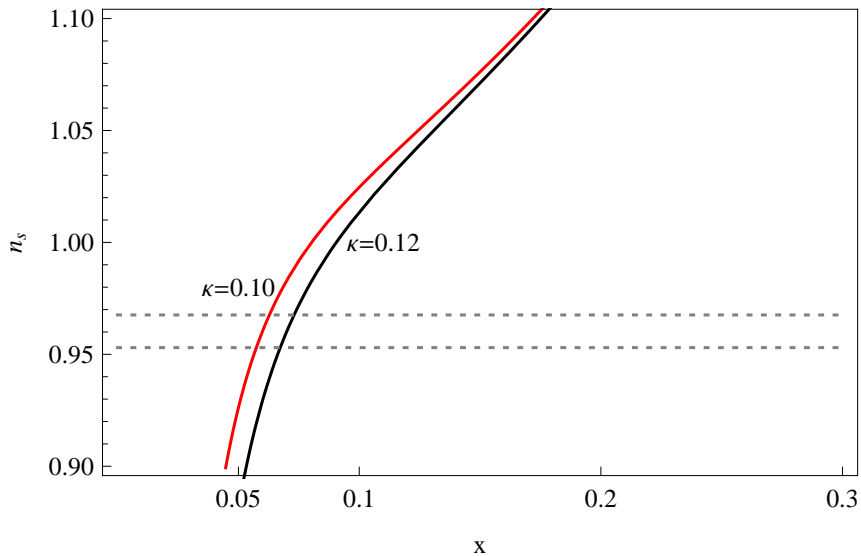}
\end{minipage}%
\caption{$n_s$ as function of $\kappa$ for initial value $x_{in}=0.3$ (red), 0.5 (black) respectively. Note that $x_{end}\geq\phi_{c}/M_{P}$. 
Dotted lines represent the uncertainty on experimental value. }
\label{nsvalue}
\end{figure}

Fig.\ref{e-folds} shows the bound on $\kappa$ for two choices of $x_{in}=0.3$ (red curve) and 0.5 (black curve)  respectively. 
Note that $x$ is constrained from slow roll condition $\mid\eta\mid<1$. 
Given the range shown in Eq.(\ref{constraints}) for $\kappa$,
$x$ should be below unity, 
which implies that large field inflation is excluded under our assumption.
Fig.\ref{e-folds} shows that $\kappa\sim 0.1$ for $N\sim 50-60$. 

In Fig.\ref{nsvalue} we show how $n_s$ changes for two typical choices of 
$\kappa$ subtracted from Fig. \ref{e-folds}.
It clearly indicates that  for observed value of $n_s$, $r$ is of order $\sim 10^{-5}$.
The simplest hybrid inflation can provide large e-folds number $N\sim 50-60$ and small $r\sim 10^{-5}$.  
Nevertheless, large $N$ and large $r >> 10^{-5}$ can not be induced at the same time.
In the next section, 
we focus on reheating after (or during) inflation.

\section{Reheating in High-Scale SUSY}
When inflation ends the conversion of energy to the MSSM matters 
from the inflaton begins immediately.
The efficiency of energy transfer depends on how inflaton is coupled to the MSSM matters,
the magnitude of their couplings,  and the SUSY mass spectrum. 
In general, the ways of energy transfer include the perturbative and non-perturbative decay of inflaton \footnote{For reviews, see, e.g., \cite{0507632, 1001.0993}.}.
The later way is known as preheating \cite{9405187,9704452, 9711360}. 
The conditions between these two ways of energy transfer are rather different.
In the later case, parameter resonance requires a quartic interaction term $\sim g^{2}\phi^{2}\chi^{2}$ with large magnitude of $g$.
This only happens if one allows renormalizable superpotential term of mass dimension 4 \cite{0603244},
\begin{eqnarray}{\label{superpotential}}
R_{\Phi}=+1: \Phi\mathbf{H}_{u}\mathbf{H}_{d};~~~~~~~~~  
R_{\Phi}=-1: \Phi\mathbf{H}_{u}\mathbf{L},
\end{eqnarray}
where $\Phi$ is the inflaton superfield, and $\mathbf{H}_{u,d}$ are Higgs doublet superfields.
$R_{\Phi}$ denotes the $R$-parity of inflaton,
which is useful to keep the dark matter stable.
In contrast, in the SM the inflaton couples to SM chiral fermions and gauge bosons 
in terms of non-renormalizable interactions of mass dimension 5.
Quartic term above doesn't exist in the SM, and therefore the way of energy transfer in the SM is perturbative decay. 
In what follows, we consider these two ways separately.

\subsection{Perturbative Decay}
As briefly mentioned above, perturbative decay happens either when there is no renormalizable interaction in Eq.(\ref{superpotential}) or the quartic coupling constant is tiny.
Instead, the inflaton only decays to SM matters via five-dimensional operators such as 
\begin{eqnarray}{\label{operators}}
\{\frac{\phi}{M} F_{\mu\nu}F^{\mu\nu},~~~~\frac{\phi}{M}\phi(H\bar{q}_{L})q_{R},~~~~ \cdots \}.
\end{eqnarray}
Here $F_{\mu\nu}$s refer to strengths of SM gauge fields,
$q$s refer to SM fermions and $M$ represents the mass scale 
appearing in the five-dimensional operators.
The plasma will be MSSM-like if the reheating temperature is larger than the typical scale of superpartner mass, $m_{0}$. 
Otherwise, the plasma is actually SM-like.

Now we calculate the reheating temperature.
We organize the decay width of inflaton to SM particles as,
\begin{eqnarray}{\label{width}}
\varGamma_{d}\equiv\frac{\lambda}{16\pi^{2}}\left(\frac{m_{\phi}}{M}\right)^{2}m_{\phi}.
\end{eqnarray}
We simply take $M=M_{P}$ but leave $\lambda$ as a free parameter.
The thermal equilibrium of relativistic plasma is dominated by $\varGamma$, 
the rate for SM inelastic scatterings of $2\rightarrow3$ processes \cite{perturbative} .
$\varGamma$ is related to $\varGamma_{d}$ as ,
\begin{eqnarray}{\label{rate}}
\varGamma\sim \alpha^{3}\left(\frac{M_{P}}{m_{\phi}}\right)\varGamma_{d},
\end{eqnarray}
where $\alpha\sim 1/30$ is the SM fine structure constant.
Therefore the reheating temperature $T_{R}$ for the perturbative decay isn't equal to the conventional one, $T_{rth}$ defined as $T_{rth}\simeq 0.3\times\left(\frac{100}{g_{*}}\right)^{1/4} \sqrt{\varGamma_{d}M_{P}}$, $g_{*}$ denotes the number of relativistic number.
Instead, in terms of Eq.(\ref{rate}) we have
\begin{eqnarray}{\label{temperature}}
T_{R}\simeq \alpha^{3/2}\cdot \left(\frac{2\pi N_{c}}{0.09}\frac{M_{P}}{m_{\phi}}\right)^{1/2}\cdot
\left(\frac{g^{susy}_{*}}{g^{sm}_{*}}\right)^{1/4}\cdot T_{rth}
\simeq 0.01\cdot \left(\frac{g^{susy}_{*}}{g^{sm}_{*}}\right)^{1/4} \cdot \left(\frac{\lambda^{1/2}}{4\pi}\right)\cdot m_{\phi},
\end{eqnarray}
where $N_c$ denote the quantum numbers of SM gauge groups.
Note that Eq.(\ref{temperature}) is valid for $m_{0}<T_{R}$,
which implies that the reheating temperature  can serve as the upper bound on $m_{0}$.

\begin{figure}
\centering
\begin{minipage}[b]{0.7\textwidth}
\centering
\includegraphics[width=4in]{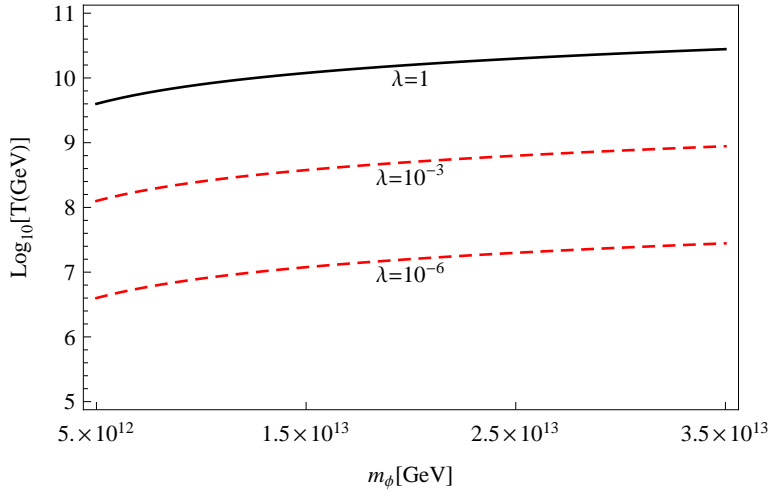}
\end{minipage}%
\caption{Reheating temperature (also the upper bound on $m_0$) as function of inflaton mass. The inflaton mass range is inferred as $\sim F/M_{P}$ up to a coefficicent.}
\label{t}
\end{figure}

In Fig.\ref{t} we show reheating temperature $T_R$ as function of inflaton mass.
Note that $\lambda$ captures the magnitude of coupling between inflaton and ``mediate " field, which also couples to the SM matter and gauge fields.
Here a few comments are in order.
(1), For $\lambda<10^{-3}$, $T_R$ is below the lower bound $\sim 1\times10^{9}$ GeV required by  thermal leptogenesis in the whole range of $m_{\phi}$.
(2), Since $T_R$ is the upper bound on superpartner mass spectrum $m_0$,
one finds that $m_{0}$ is upper bounded as $1$ TeV $<<m_{0}<10^{7}$ TeV for the case of perturbative decay.
(3), As $T_R$ is far below the gravitino mass of Eq.(\ref{gravitinomass}),
there is no overproduction problem of gravitino in high-scale SUSY.
Superheavy gravitino mass of order $\sim 10^{14}$ GeV kinetically blocks its production in the thermal bath.

\subsection{Non-perturbative Decay }
If it admits renormalizable superpotential Eq.(\ref{superpotential}), 
there exists quartic interaction between inflaton and its decay products $\chi_i$,
Non-perturbative decay can happen in wide range of parameter space for 
potential of type\footnote{Assuming inflaton and MSSM matters share the same origin of SUSY breaking, inflaton mass is dynamical induced by SUSY breaking. 
In this sense, the mass term $m^{2}_{\phi}\phi^{2}$ is a soft SUSY-breaking term other than 
arises from SUSY tree-level mass superpotential $\sim m\Phi\Phi$.
Consequently, there is no cubic interaction $\phi\chi^{2}$ in compared with earlier discussions in \cite{0603244, 0512227}.},
\begin{eqnarray}{\label{nonpp}}
V(\phi, \chi)= \frac{1}{2} m^{2}_{\phi}\phi^{2}+g^{2}\phi^{2}\chi^{2}_{i}+m^{2}_{\chi_{i}}\chi^{2}_{i},
\end{eqnarray}
where the inflaton mass term is included and $m_{\chi_{i}}$ is the mass for $\chi_i$.
We would like to mention that $m_{\chi_{i}}$ include soft SUSY breaking contribution of order $\sim m_{0}$ and dynamical mass $\sim \lambda^{1/2} \left<\varphi\right>$ induced by VEV of flat direction $\varphi$ \cite{negative} through the quartic interaction \cite{ 0512227},
\begin{eqnarray}{\label{flat}}
V(\chi, \varphi)=\lambda \chi_{i}^{2}\varphi^{2},
\end{eqnarray}
where $\lambda$ is the quartic coupling constant.
The magnitude of $\left<\varphi\right>$ is determined by the self-interaction potential for flat direction $V(\varphi)$.

Now we consider the potential for flat direction.
$V(\varphi)$ includes soft breaking mass, Hubble parameter induced term
and high dimensional operators, 
\begin{eqnarray}{\label{varphi}}
V(\varphi)\simeq (m^{2}_{0}+c_{H}H^{2})\varphi^{2}+c_{6}\frac{\varphi^{6}}{M^{4}}+\cdots,
\end{eqnarray}
where $c_{H}$ is real coefficient.
Since $m_0$ isn't far beneath the Hubble constant $H\sim m_{\phi}$ at the beginning of inflation,
VEV $\left<\varphi\right>$ depends on the sign of  $c_{H}$, 
which can be either positive  or negative \cite{negative,positive}.  
In particular, $\left<\varphi\right>=0$ for the case of either positive $c_H$ or negative $c_{H}$ but with $\mid c_{H}\mid <<1$.
It implies that SM gauge symmetry is unbroken during the whole history of early universe.
On the other hand, $\left<\varphi\right>\neq 0$ for negative $c_{H}$ but with $\mid c_{H}\mid >1$. It implies that SM gauge symmetry is broken in the eary universe,
with gauge boson mass of order $\left<\varphi\right>$, 
then restored after the epoch of reheating.

The potentials we define in Eq.(\ref{nonpp}) to Eq.(\ref{flat}) are rather general,
which can be applied to both cases in Eq.(\ref{superpotential}).
To discuss the condition for parameter resonance,
one starts with the modified Klein-Gordon equation for Fourie modes $\chi$.
Whether WKB approximation is viable for the study can be analyzed in term of a quantity $R$ defined as \cite{0507632},
\begin{eqnarray}{\label{R}}
R\equiv \frac{\dot{\omega}_{k}}{\omega_{k}^{2}},~~~~~~~~\omega^{2}=k^{2}/a^{2}+g^{2}\left<\phi(t)\right>^{2},
\end{eqnarray}
where dot refers to derivative over time and $\omega$ is the frequency, 
with $a$ the expansion factor and $k$ the momentum.  
If $\mid R\mid<<1$,  the WKB approximation is valid, 
the produced particle number of $\chi$ doesn't grow in this case. 
If $\mid R\mid>1$ instead,  the WKB approximation isn't valid,
which leads to significant production of $\chi$.
In long wavelengths limit, this constraint is given by 
\footnote{There is a coefficient of order one in front of $m_{0}$ for either $R_{\Phi}=1$ or $R_{\Phi}=-1$. Here we simply take it equal to unity.}, 
\begin{eqnarray}{\label{c1}}
m^{2}_{\chi_{i}}\simeq m^{2}_{0}+\lambda\left<\varphi\right>^{2}< g^{2}\left<\phi(t)\right>^{2},
\end{eqnarray}
where we have used Eq.(\ref{nonpp}) and Eq.(\ref{flat}). 
Moreover, in order to keep that the parameter resonance isn't spoiled by expansion, 
an additional constraint must be imposed,
\begin{eqnarray}{\label{c2}}
q\equiv g^{2}\bar{\phi}^{2}(t)/4m_{\phi}^{2} >>1,
\end{eqnarray}
where $\bar{\phi}(t)\sim M_{P}$ refers to the amplitude of inflaton oscillations.
For more details, we refer to reader to \cite{0507632} and references therein.

\begin{figure}
\centering
\begin{minipage}[b]{0.5\textwidth}
\centering
\includegraphics[width=3.2in]{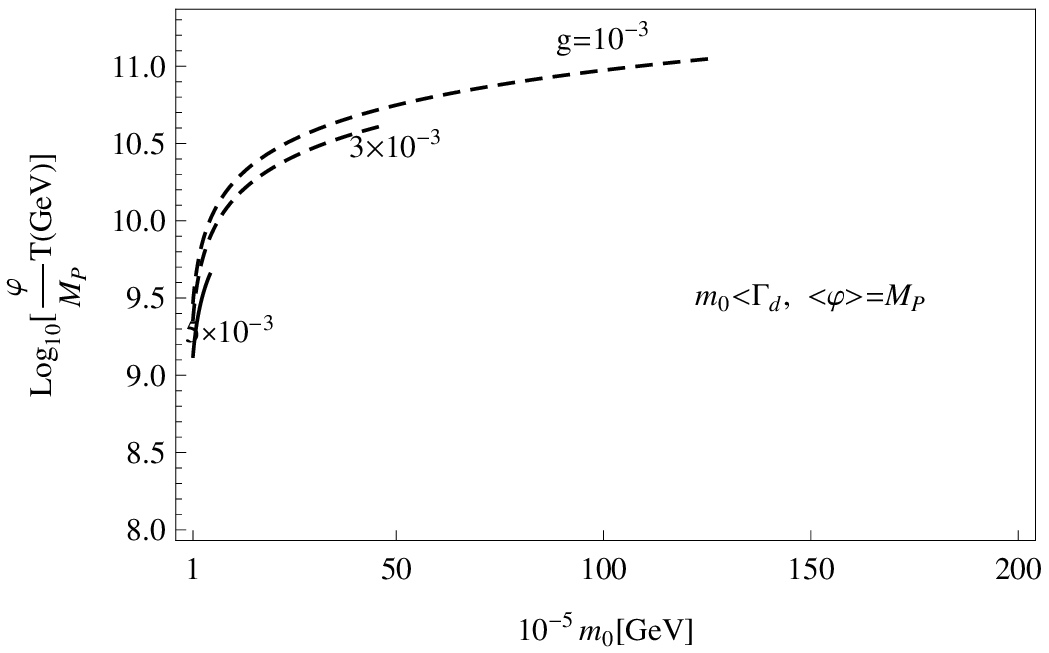}
\end{minipage}%
\centering
\begin{minipage}[b]{0.5\textwidth}
\centering
\includegraphics[width=3.2in]{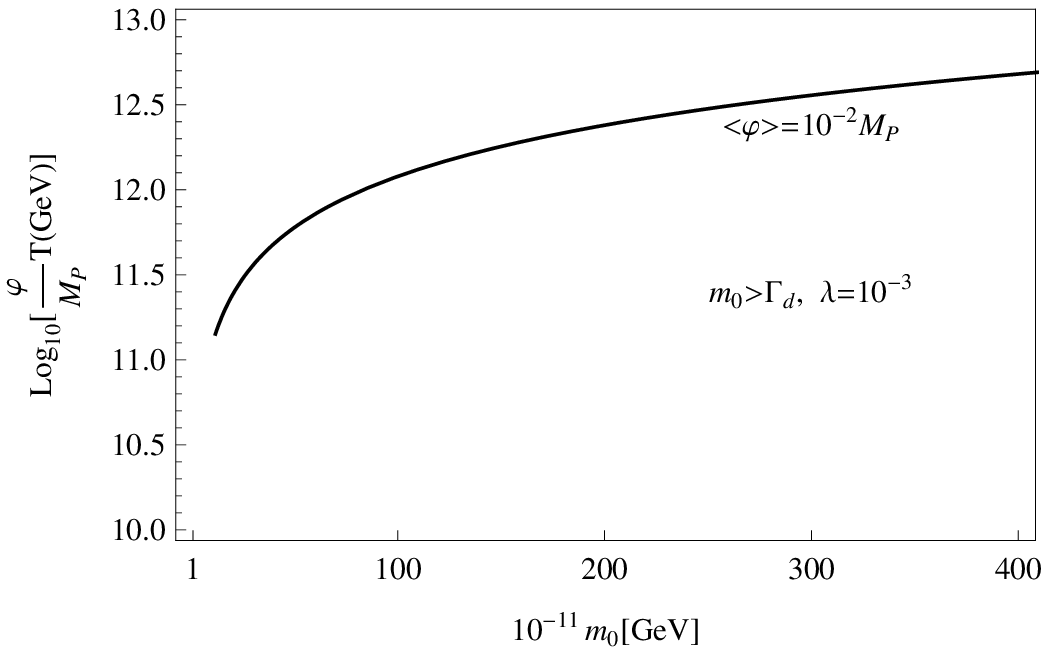}
\end{minipage}%
\caption{Reheating temperature as function of $m_0$ for $m_{\phi}=10^{13}$ GeV and $\left<\varphi\right>\neq 0$.
($\mathbf{Left}$: $m_{0}<\varGamma_{d}$ with $\left<\varphi\right>=M_{P}$, $\mathbf{Right}$: $m_{0}>\varGamma_{d}$ with $\lambda=10^{-3}$ and  $\left<\varphi\right>=0.01	M_{P}$).
In the $\mathbf{Right}$ panel, 
$\left<\varphi\right>\geq 0.1M_{P}$ is excluded by condition Eq.(\ref{c1}) for $\lambda\geq 10^{-3}$.
The bounds on $g$ and $m_0$ are explained in the text.  }
\label{nonzerov}
\end{figure}

Fig.\ref{nonzerov} shows reheating temperature for the case in which $\left<\varphi\right>\neq 0$.
In this case, SM gauge bosons are massive and the rate for thermal equilibrium $\varGamma$ \cite{0512227}  depends on its magnitude relative to $m_0$.
For $m_{0}<\varGamma_{d}$, 
$T_R$ depends on both $g$ and $\left<\varphi\right>$,
whereas it mainly depends on $\left<\varphi\right>$  for $m_{0}>\varGamma_{d}$.
In this figure, we take $m_{\phi}=10^{13}$ GeV and $\bar{\phi}=M_{P}$.
The bounds on $g$ are due to a few considerations.
The first one is that condition Eq.(\ref{c2}) from parameter resonance requires $g>>10^{-5}$.
The second one is that overproduction problem \cite{gravitino}  of gravitino in high-scale SUSY can be kinematically  blocked if $g<10^{-2}$ such that non-perturbatively induced mass during parameter resonance is beneath $m_{3/2}$ 
\footnote{ 
Ref. \cite{1404.1914} provides an example how gravitino problem in high-scale SUSY is evaded in the context of mini-Split SUSY. In comparison with \cite{1404.1914}, 
gravitino mass is far heavier in this paper,
and kinematically blocking is the solution to the overproduction of gravitino.}.
The bound on $\lambda\left<\varphi\right>$ arises from  condition Eq.(\ref{c1}) which shows $\lambda\left<\varphi\right>^{2}<g^{2}M_{P}^{2}$.

The $\mathbf{Left}$ panel  in  Fig.\ref{nonzerov} shows that in the range $10^{5}$ GeV $<m_{0}<1.5\times 10^{7}$ GeV  reheating temperature $T_{R}\geq 10^{9}$ GeV in the allowed range of $g$ for $\left<\varphi\right>=M_{P}$. 
With modifying $\left<\varphi\right><M_{P}$, 
\begin{eqnarray}{\label{ntr}}
T_{R}\rightarrow \left(\frac{\left<\varphi\right>}{M_{P}}\right)^{-1}T_{R},
\end{eqnarray}
which is always above the value required by thermal leptogenesis.
The $\mathbf{Right}$ panel  in  Fig.\ref{nonzerov} shows that 
in the range $10^{11}$ GeV $\leq m_{0}<4\times 10^{13}$ GeV 
reheating temperature  $10^{11} $ GeV $\leq T_{R}\leq 10^{13}$ GeV if  $\left<\varphi\right>=10^{-2} M_{P}$,
and changes similarly  to Eq.(\ref{ntr}) for modifying $\left<\varphi\right>$.
This implies that $T_R$ is also always above the the value required by thermal leptogenesis.
Note that $\left<\varphi\right>\geq 10^{-1} M_{P}$ is excluded by condition Eq.(\ref{c1}) from parameter resonance for $\lambda\simeq 10^{-3}$.

\begin{figure}
\centering
\begin{minipage}[b]{0.7\textwidth}
\centering
\includegraphics[width=4in]{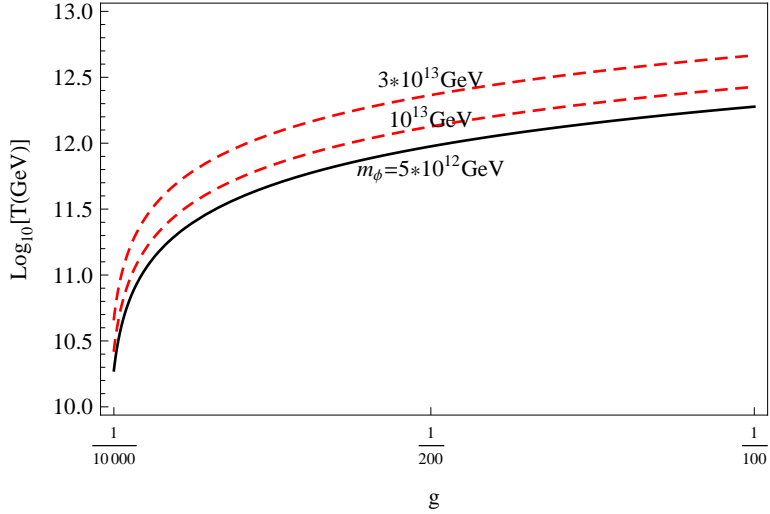}
\end{minipage}%
\caption{Reheating temperature as function of $g$ for $\left<\varphi\right>=0$,
with the definition $\varGamma_{d}\equiv g^{2}m_{\phi}/8\pi$.}
\label{zerov}
\end{figure}

Fig.\ref{zerov} shows reheating temperature for the case in which $\left<\varphi\right>=0$.
In this case, SM gauge symmetries are unbroken in the epoch of reheating.
Thermalization cannot occur before the inflaton decay has completed. 
Due to $\left<\varphi\right>=0$ the $2 \rightarrow 3$ scatterings are already efficient 
when $H\simeq \varGamma_{d}$. 
The reheat temperature in this case is given by the standard expression: $T_{R} \simeq 0.3 \sqrt{\varGamma_{d}M_{P}}$.
However, $\varGamma_{d}$ calculated via renormalizable couplings is rather different 
from Eq.(\ref{width}) calculated via non-renormazible couplings. 
This difference between Fig.\ref{t} and Fig.\ref{zerov} is obvious.
Typically, we have $T_{R}\geq 10^{10}$ GeV in non-perturbative decay into MSSM and $T_{R}\leq 10^{10}$ GeV in perturbative decay into SM.

Both Fig.\ref{zerov} and Fig.\ref{nonzerov} show that $T_R$ is above $\sim 10^{9}$ GeV 
but beneath $m_{3/2}$ in a wide range of parameter space.
Due to kinematically blocking effect this evades the overproduction of gravitino in conventional high-scale SUSY.

\section{Conclusions}
In the light of both LHC data and Plank bound on $r$, 
High-scale SUSY is more favored in compared with low-scale SUSY. 
In this paper, we discussed the implications of high-scale SUSY to the early universe.
In particular, we assumed that the inflation and visible sector share the same origin of SUSY breaking, 
and derived model independent consequences based on this assumption. 
We find that the reheating temperature  for superpartner mass spectrum above $\mathcal{O}(100)$ TeV might be beneath the value required by thermal leptogenesis if inflaton decays to its products perturbatively but above it if non-perturbatively instead.
We also observed that problem of gravitino overproduction can be evaded through kinematically blocking in a wide range of parameter space in the later way. 

As an illustration for the model building of inflation in the course of high-scale SUSY,
in section 3 we revise the simplest hybrid inflation that includes a new linear term for inflaton with coefficient proportional to $m_{3/2}$. It is shown that this term significantly affects the choices on initial condition of inflaton fields. 
We found that  with the assumption
the simplest hybrid inflation is consistent with present experimental data for $r$ of order $10^{-5}$.

Under our assumption only the dark matter is a light SUSY state with mass near the weak scale \cite{Zheng}, 
which is the target of LUX and Xenon experiments, etc. 
Hopefully, it can be addressed in the near further.

\begin{acknowledgments}
We would like to thank J.-h, Huang for discussion.
This work is supported in part by the Fundamental Research Funds for the Central Universities under Grant No.CQDXWL-2013-015.
\end{acknowledgments}

\linespread{1}

\end{document}